\documentstyle[twoside,fleqn,espcrc]{article}

\newcommand{\nc}{\newcommand}
\nc{\tb}{tan$\beta$ }
\nc{\be}{\begin{equation}}
\nc{\ee}{\end{equation}}
\nc{\nonum}{\nonumber\\}
\nc{\ba}{\begin{eqnarray}}
\nc{\ea}{\end{eqnarray}}
\nc{\mg}{$M_{\tilde{g}}$ }

\newcommand{\AmS}{{\protect\the\textfont2
  A\kern-.1667em\lower.5ex\hbox{M}\kern-.125emS}}

\title{Aspects of Supersymmetry at LEP II}

\author{D. Matalliotakis and H.P. Nilles\address{Physik Department,
        Technische Universit\"at M\"unchen, \\
        D-85747 Garching, Germany, \\
        and \\
        Max Planck Institut f\"ur Physik, Werner Heisenberg Institut,\\
        D-80805 M\"unchen, Germany}%
        \thanks{Talk given by H.P. Nilles at the Zeuthen Workshop on
          Elementary Particle Theory--Physics at LEP200 and Beyond,
          Teupitz/Brandenburg, Germany, 10--15 April, 1994. To appear in
          the proceedings edited by T. Riemann and J. Bl\"umlein.
          This work is supported by
          DAAD, DFG and EU grants SC1-CT91-0729 and
          SC1-CT92-0789}}

\begin{document}

\begin{abstract}
We give a short introduction to the concept of low energy supersymmetric
models and their phenomenological predictions. In view of the future
LEP II results we have to parametrize these predictions in a model
independent way without too many simplifying assumptions. The study must
therefore include models with {\it non--universal} soft supersymmetry
breaking terms.  We show that such non--universalities might have
significant influence on the properties of these models.
\end{abstract}

\maketitle

\section{INTRODUCTION}

The future results of LEP II will most probably play a decisive role in the
search for signals of low energy supersymmetry. Wide regions of parameter
space of the supersymmetric extensions of the $SU(3)\times SU(2)\times
U(1)$ standard model can be tested. But there is still some time before
LEP II comes into operation. Meanwhile we shall learn more from LEP I, the
Tevatron and maybe other experiments. In any case the time for a thorough
scan of parameter space will soon come. We can no longer make
simplified assumptions and just consider simple models, even if they
might be justified from a theoretical or aesthetic point of view. The
quantitative phase in the search for supersymmetry has begun. In these
lectures we shall discuss some aspects of this general search and
especially concentrate on the implications of a possible non--universality
of the soft supersymmetry breaking terms. Such non--universalities
might play a crucial role in the SUSY--predictions that are relevant
for LEP II.

\section{LOW--ENERGY SUPERSYMMETRY}

The motivation for the study of the supersymmetric extensions of the
standard model is the desire to understand the stability of $M_W$, the
weak scale represented by the mass of the intermediate gauge bosons.
Why is $M_W$ so small compared to $M_X$ (a possible grand unified
scale of order $10^{16}$GeV) or $M_{\rm Planck}$ ($\sim 10^{19}$GeV)?
This puzzle is connected to the question of quadratic divergences that
might destabilize the mass of the Higgs boson in a non--supersymmetric
theory.

Supersymmetric extensions of the $SU(3)\times SU(2)\times
U(1)$ standard model contain in addition to the gauge bosons, fermions
and Higgs bosons new particles to complete vector and chiral supermultiplets.
These are the gauginos, scalar partners of quarks and leptons
as well as fermionic partners of the Higgs bosons. A second Higgs
supermultiplet has to be introduced to avoid problems with gauge
anomalies and allow a mass mechanism for all quarks and leptons. Space does
not permit us to give a detailed discussion of the constructions of such
models. For comprehensive reviews see ref.
\cite{nil,sec}. They include a
discussion of the two major questions raised in these models:
baryon number violation via dimension--4--operators and the $\mu$--problem
of the Higgs masses. The former leads to the introduction of a new symmetry
(e.g. R--parity) while a resolution of the other seems to imply new
structure at higher scales.

The construction of realistic models requires the incorporation of
local supersymmetry (i.e. supergravity). This framework allows a
satisfactory generation of scalar masses and provides a mechanism
for (continuous) R--symmetry breakdown to allow for non--vanishing
gaugino masses. Finally, in contrast to the globally supersymmetric
case, broken {\it local} supersymmetry is consistent with a
vanishing cosmological constant. These models consist of two sectors,
very weakly coupled to each other only through gravitational
interactions. The so--called {\it observable} sector consists of the
particles of the standard model including the supersymmetric
partners. A second {\it hidden} sector is responsible for spontaneous
supersymmetry breakdown at an intermediate scale
$M_S\sim 10^{11}$GeV. Through the gravitational coupling of the two
sectors on obtains mass splittings
$m_{3/2}\sim M_S^2/M_{\rm Pl}\sim 1$TeV in the observable sector.
The mass splittings in turn give rise to an induced breakdown of
$SU(2)\times U(1)$ now linked to the breakdown of supersymmetry.
The low energy effective theory can then be parametrized by a globally
supersymmetric theory with soft SUSY breaking terms that originate
 from the spontaneous breakdown of supergravity. These include scalar
masses for the partners of quarks, leptons and Higgses, gaugino
masses as well as SUSY breaking trilinear interaction terms. The order
of magnitude of these soft terms is set by the gravitino mass
$m_{3/2}$ of order of a TeV. The magnitude of the soft terms
determines the spectrum of the supersymmetric partners and present
experimental limits place restrictions on the values of these soft
parameters. We are eagerly waiting for LEP II to detect
supersymmetric particles or to improve these limits.
Before this machine comes into operation we have to settle with more
indirect searches like the precision tests at LEP I.

\section{SUSY--GUTS AND SOFT PARAMETERS}

The apparent
 ``success" of supersymmetric grand unified theories (SUSY--GUT's)
might be taken as a hint for a possible presence of supersymmetric
particles in the TeV--region or below. It at least gives us a justification
to consider these theories seriously and work out the
phenomenological consequences of the assumptions that concern
gauge coupling unification, Yukawa coupling unification and the
requirement of radiative breakdown of weak $SU(2)\times U(1)$.
There are, of course, many uncertainties in such theories that are
related to its properties at the grand unified scale. These include
threshold effects due to heavy particles as well as the question
concerning the actual value of the parameters to be put in at that
grand scale. This would include also the soft SUSY breaking terms. They
are signals from the underlying supergravity or superstring theory
and depend strongly on the mechanism of supersymmetry breakdown.
At the moment we can only speculate about the spectrum of these soft
terms and should therefore work out the phenomenological
consequences for a general set of parameters.

The simplest choice would correspond to universal soft terms at the
GUT--scale. Usually one assumes universality of scalar masses
$m_{0}^{2}$, gaugino masses $M_{1/2}$ and A--parameters. The absence of
flavour changing neutral currents puts constraints on the mass
difference of squarks with the same charge.  Universality of
squark masses can be achieved through the choice of minimal
kinetic terms in an underlying supergravity theory. It would imply that
the non--gauge interactions have universal coupling strengths to
all matter fields. In any case the choice of universal
terms is an ad hoc assumption.
Other choices might be motivated equally well  and deserve further
attention given the precision with which we can determine the gauge coupling
constants.

There also exist theoretical arguments for a non--universality of the mass
terms. In many supergravity models the Higgs superfields belong to
 a different sector than the matter superfields containing quarks
and leptons. The Higgs multiplets might reside in sectors with
N=2 supersymmetry and are more closely related to the gauge multiplets
than to the N=1 chiral matter superfields. Since there are no restrictions on
the Higgs masses from flavour changing neutral currents, one might consider the
Higgs--mass as a separate input parameter that might differ from
$m_{0}^{2}$ of the matter fields. Moreover, explicit examples of
orbifold string theories indicate the possibility of
non--universal soft terms \cite{str} and the same is true in general
supergravity
theories \cite{sgr}.

A second related question concerns the actual scale at which the
(universal) terms are defined. In the underlying theory the
fundamental scale (e.g. the Planck scale $M_{Pl}$) might differ
 from $M_{X}$. Even in a framework of universal soft terms (most
naturally chosen at $M_{Pl}$) one would encounter non--universal
behaviour at $M_{X}$ \cite{pom}. Heavy threshold effects might strongly
influence the spectrum of the soft terms at the very high scales.

Given this situation it seems to be mandatory in a discussion of SUSY--GUT's
to work out explicitly the consequences of non--universal soft terms.
At present we know a lot about SUSY--GUT's with universal terms
\cite{op,copw1,bbo,abs,kkrw,copw2,gp}.
If we include the assumption of Yukawa coupling unification and radiative
electroweak symmetry breaking, we are in many cases led to a very
restrictive model with some definite predictions. We have then to
investigate the stability of these predictions in the absence of
universal mass terms. These are the questions we want to study in the
next section. Certain aspects of non--universal breaking terms have already
been addressed in ref. \cite{pom,nu}.

\section{IMPLICATIONS OF NON-UNIVERSALITY OF SOFT TERMS}

In our analysis  \cite{mn}
we assume the Minimal Supersymmetric Standard Model (MSSM) particle
spectrum below the unification scale and require:\\
i) Exact gauge coupling unification at $M_X$.\\
ii) Third generation Yukawa coupling unification: We will examine in turn
the large and small \tb regimes (tan$\beta=\frac{\upsilon_{2}}{\upsilon_{1}}$
is the ratio of the vevs of the two Higgs doublets),
requiring in the first case that all
three  Yukawas unify (SO(10)--like unification), while in the second case
that only the bottom and tau Yukawa couplings unify (SU(5)--like
unification). Again exact unification at the scale $M_X$ is assumed.\\
iii) Radiative electroweak breaking: Starting with a symmetric theory at
the unification scale, we require a proper breakdown of the
SU(2)$\times$U(1) electroweak symmetry induced through radiative corrections
to the Higgs sector parameters.

In imposing gauge and Yukawa coupling unification we use the
$\overline{MS}$--values \cite{lp3}
\ba
&&sin^{2}\theta_{W} = 0.2324 -0.92\cdot10^{-7}\cdot\nonum
&&\ \ \ \ \left(M_t^2-(143GeV)^2\right)GeV^{-2}\pm 0.0003\ ,\nonum
&&\alpha_{em}^{-1} = 127.9\pm0.1\ ,
\ea
of the electroweak couplings at $M_{Z}$
to determine $M_{X}$ and $\alpha_{G}$,
the unification scale and the gauge coupling at this scale
respectively. The pole--mass value $M_{\tau}$=1.78GeV \cite{pd} of the
tau lepton is used to determine the common value
$h_{b}(M_{X})=h_{\tau}(M_{X})$ of the b-- and $\tau$--Yukawa
couplings at $M_{X}$ respectively. In SO(10)--like unification
schemes no further low energy input is needed to determine
$h_{t}(M_{X})$, the top Yukawa coupling at $M_{X}$,
since it unifies with the other two. When we require SU(5)--like
unification however, we use the bottom quark pole--mass
4.7GeV$\le{M_{b}}\le{5.2}$GeV \cite{pd} to fix the value of $h_{t}$
at $M_{X}$.
The value of \tb is also an input in our analysis. Unification
gives us a prediction for $\alpha_{3}(M_Z)$, the strong gauge
coupling at $M_{Z}$, and $M_{t}$, the top quark pole--mass
given a value of tan$\beta$.
In SO(10)-like scenarios we have in addition a prediction of the
bottom quark mass\footnote{In these scenarios one could trade
the prediction of $M_{b}$ for that of the \tb value.}. In the determination
of the running masses from the pole--mass data we neglect QED
corrections, while taking QCD corrections into account at the two--loop
level \cite{a2}. Renormalization below $M_{Z}$ is carried out via two--loop
QED and three--loop QCD renormalization group equations (RGE's) and
the value of the strong gauge coupling used, is the one predicted from
unification in each particular case that we study.

To address the question of radiative electroweak symmetry breaking we start
by assuming that low energy supersymmetry originates from
a spontaneously broken supergravity theory. This theory
is spontaneously broken by some unknown mechanism at the Planck or
string scale, and the effect of this breakdown is
parameterized by adding to the globally supersymmetric effective
 Lagrangian a series of soft breaking terms. This supersymmetry
breaking part of the Lagrangian has the form:
\ba
-{\cal L}_{soft} &=& m_{Q}^2\left|Q\right|^{2}+m_{U}^2\left|U\right|^{2}+
m_{D}^2\left|D\right|^{2}\nonum
&&{}+m_{L}^2\left|L\right|^{2}+
m_{E}^2\left|E\right|^{2}\nonum
&&{}+m_{H_{1}}^2\left|H_{1}\right|^{2}+
m_{H_{2}}^2\left|H_{2}\right|^{2}\nonum
&&{}+[h_{t}A_{t}QH_{2}\bar{U}+h_{b}A_{b}QH_{1}\bar{D}\nonum
&&{}+h_{\tau}A_{\tau}LH_{1}\bar{E}+\mu BH_{1}H{_2}+h.c]\nonum
&&{}+\frac{1}{2}M_{3}\bar{\lambda}_{3}\lambda_{3}+
\frac{1}{2}M_{2}\bar{\lambda}_{2}\lambda_{2}\nonum
&&{}+\frac{1}{2}M_{1}\bar{\lambda}_{1}\lambda_{1}\ ,
\label{eq:st}
\ea
where $m_{i}$, $A_{i}$, $B$, $M_{i}$ are a set of parameters with mass
dimension one, $h_{i}$ are the Yukawa couplings and $\mu$
is the coefficient of the Higgs mixing term in the superpotential.
All fields appearing in (\ref{eq:st}) are scalar, except for the gauginos
which are Majorana fermions. Generation indices have been
suppressed.

 From this Lagrangian one can extract the potential for the
neutral components of the Higgs fields:
\ba
V&=&m_{1}^{2}\left|H_{1}^{0}\right|^{2}+
m_{2}^{2}\left|H_{2}^{0}\right|^{2}-
m_{3}^{2}(H_{1}^{0}H_{2}^{0}+c.c)\nonum
&& {}+\frac{\lambda_{1}}{2}\left|H_{1}^{0}\right|^{4}+
\frac{\lambda_{2}}{2}\left|H_{2}^{0}\right|^{4}\nonum
&&{}+(\lambda_{3}+\lambda_{4})\left|H_{1}^{0}\right|^{2}
\left|H_{2}^{0}\right|^{2}\ ,
\label{eq:hp}
\ea
where $m_{1}^{2}\equiv{m_{H_{1}}^{2}+\mu^{2}}$,
$m_{2}^{2}\equiv{m_{H_{2}}^{2}+\mu^{2}}$,
and we take $m_{3}^{2}\equiv{B\mu}$ to be positive.
At scales where our theory is (softly broken)
supersymmetric the quartic couplings $\lambda_{i}$
satisfy the relations
\ba
&&\lambda_{1}=\lambda_{2}=\frac{1}{4}(g_{1}^{2}+g_{2}^{2})\ ,\nonum
&&\lambda_{3}=\frac{1}{4}(g_{2}^{2}-g_{1}^{2})\ ,\ \ \
\lambda_{4}=-\frac{1}{2}g_{2}^{2}\ ,
\ea
with $g_{1}$, $g_{2}$ being the U(1), SU(2) gauge couplings of
the Standard Model respectively.

The parameters appearing in the Higgs potential (\ref{eq:hp})
are evaluated at the unification scale and then renormalized down to the
electroweak scale. The simplest choice of boundary conditions is the
universal one
\ba
&&m_{H_{1}}^{2}=m_{H_{2}}^{2}=m_{Q}^{2}=\cdots=
m_{E}^{2}=m_{0}^{2}\ ,\nonum
&&{A_{t}=A_{b}=A_{\tau}=A}\ ,\nonum
&&M_{3}=M_{2}=M_{1}=M_{1/2}\ ,
\ea
already extensively studied in the literature.

The aim of the present analysis is to investigate the
phenomenological implications of non--universal soft
breaking terms at the unification scale. We will mainly concentrate on
two examples of non--universality, namely:\\
i) Disentangling the Higgs soft masses from the rest of the
scalars:
\ba
&&m_{H_{1}}^{2}=m_{H_{2}}^{2}=m_{H}^{2}\ ,\nonum
&&m_{Q}^{2}=\cdots  = m_{E}^{2}=m_{0}^{2}
\label{eq:hibo}
\ea
at $M_{X}$, with $m_{H}$ being independent of $m_{0}$.\\
ii) Relaxing the universality within the Higgs sector:
\ba
&&m_{H_{1}}^{2}\neq m_{H_{2}}^{2}\ .
\ea
We in addition investigated the effect of non--universal gaugino masses
and we will also briefly comment on the effect of lifting the
universality between up-- and down--type squarks.

We follow an up--down approach, renormalizing parameters from
the unification scale down to low energies. Two--loop RGE's
are used for the gauge and Yukawa couplings, one--loop
for the soft mass parameters. In examining the breaking of the
electroweak symmetry we minimize at $M_{Z}$ the
renormalization--group improved tree--level Higgs potential,
and take into account the corrections coming from the mass
splitting in the supermultiplets by decoupling heavy
superpartners below some common scale $M_{S}$. The most
sizeable corrections to the Higgs potential parameters
come from the mass splitting in the quark and gluon
supermultiplets. We therefore decouple the squarks and the gluino at
the scale $M_{S}$ and use below this scale the RGE's given
in ref. \cite{cha}, with the appropriate matching conditions
for the Higgs potential parameters at the threshold \cite{cha,op}.
$M_{S}$ is chosen so as to minimize these threshold corrections
and we
require stability of our results against reasonable variations
of $M_{S}$ in the range of the squark masses. Between $M_{X}$
and $M_{S}$ the well--known MSSM RGE's are used (for easy
reference see e.g. the appendix in ref. \cite{op}).

In scanning the parameter space of the soft masses, we have excluded
solutions in which the squarks are too heavy.
For practical purposes a limit of $\sim$2.5TeV has been imposed
on the squark masses.
We also impose the experimental lower bounds on the sparticle
spectrum, the most relevant being a lower bound of approximately
110GeV on the gluino and 45GeV on the lightest stop.

A crucial parameter of the MSSM is tan$\beta$. As we have already
mentioned, the requirement of gauge and b--$\tau$ Yukawa
coupling unification gives us a prediction of the top quark
pole--mass as a function of \tb \cite{cpw,lp2}.
For the top quark to be lighter
than 200GeV (and indeed in the range recently reported by CDF \cite{cdf})
two disjoined regimes for \tb are allowed: a low regime
with \tb of the order of one and a high regime with \tb
between approximately 35 and 60. One should
note that this prediction ignores corrections to the bottom
and tau masses induced by heavy sparticle loops, which can
be quite important when \tb is large \cite{hrs1}. We will further
comment on this point later on.

We performed our analysis for both \tb
regimes. When \tb is large it can account for the mass
hierarchy between the top and the bottom quark so the corresponding
Yukawa couplings have similar values. It is possible then
to assume that also $h_{t}$ unifies with $h_{b}$ and $h_{\tau}$,
a scenario naturally realizable in the
SO(10) unification scheme. On the other hand
this similarity of the top and bottom Yukawa coupling values
makes the breaking of the electroweak symmetry through radiative
corrections difficult to achieve when \tb is large. The reason
for this is that the masses of the two Higgses evolve very
similarly when there is no difference between $h_{t}$ and $h_{b}$,
so the condition $m_{1}^{2}m_{2}^{2}<m_{3}^{4}$
for destabilizing the tree--level Higgs potential at the
origin can only be fulfilled for some carefully chosen regions
in the space of soft parameters when universality at
$M_{X}$ is assumed. This is not the case in the low \tb
regime where $h_{t}\ll{h_{b}}$
so the product $m_{1}^{2}m_{2}^{2}$ can easily be driven
negative in large regions of the parameter space, even if one
starts with the two Higgs bosons degenerate at $M_{X}$. Therefore it
 should be noted that although non--universalities can exist
in both scenarios, their role in the large \tb scheme might be crucial
for a more natural realization of radiative electroweak breaking,
since they can induce the required hierarchy of the Higgs masses.

We start the presentation of our results with the large \tb case.
To be concrete we choose the specific value of tan$\beta$=55, but
the effects are qualitatively the same for the whole range,
for which SO(10)--like unification is phenomenologically
viable. For this value of \tb the unification prediction for the strong
gauge coupling is $\alpha_{3}(M_{Z})\simeq$0.130, for the
top quark pole--mass $M_{t}\simeq$192GeV and for the bottom pole--mass
$M_{b}\simeq$5.5GeV. These predictions are obtained if the
supersymmetric RGE's are integrated all the way down to $M_{Z}$,
while slightly lower values for $\alpha_{3}(M_{Z})$ and $M_{t}$
($\sim$0.126 and $\sim$186GeV respectively) are predicted if
one decouples the sparticles at an effective scale $T_{SUSY}$ larger
than $M_{Z}$ \cite{lp1,cpw}. Note that the prediction for $M_{b}$ is
somehow larger than the upper bound quoted for this quantity
($\sim$5.2GeV). One should however be cautious, since
for large  \tb the corrections to the
bottom mass coming from heavy sparticle loops can be very big depending
on the spectrum of the superpartners. We will address this issue
in our discussion of the radiative electroweak symmetry breaking.

The dominant features observed in our analysis can be read off from figures
1 through 4. In each of these figures three different kinds of
regions can be identified, corresponding to the following three cases
studied:\\
{\bf Case A} -- Universal soft breaking terms: Depicted by the shaded regions
enclosed in dotted lines.\\
{\bf Case B} -- Independent Higgs soft parameter: A Higgs soft parameter
$m_{H}$
independent of $m_{0}$ has been introduced and the boundary
conditions (\ref{eq:hibo}) are imposed at $M_{X}$. The solution
points lie within the solid lines.\\
{\bf Case C} -- Non--degenerate Higgs doublets: The solution points enclosed
by the dashed lines correspond to the following boundary conditions
at the unification scale:
\ba
&&m_{H_{1}}=m_{0}\ ,\nonum
&&m_{H_{2}}=0.8{m_{0}}\ ,
\ea
where $m_{0}$ is the common mass parameter for squarks and sleptons
at $M_{X}$.

Our main conclusion from the study of non--universalities in the large
\tb regime is that radiative electroweak breaking and grand unification
can be realized much more naturally, if one starts with non--universal
soft masses within the Higgs sector. Aspects of this conclusion are
present in all four figures presented. In Fig. 1 we show the
$m_{0}$--$M_{1/2}$ plane. In the universal case it is known that proper
symmetry breaking occurs only when  $m_{0}$ is smaller or at most of
the order of $M_{1/2}$. A somehow
simplified way to understand this is the following: When
$h_{b}\sim{h_{t}}$ the two main features that differentiate the running
of the two Higgs mass parameters $m_{H_{1}}^{2}$, $m_{H_{2}}^{2}$ are the
asymmetric appearance of the up-- and down--type squark mass parameters
$m_{U}^{2}$, $m_{D}^{2}$ in their RGE's,
as well as the contribution of the tau Yukawa coupling to the equation for
$m_{H_{1}}^{2}$ only. Due to the larger hypercharge of the right--handed
top
compared to the right--handed bottom, large
values of $M_{1/2}$ tend to
increase $m_{U}^{2}$ with respect to $m_{D}^{2}$ and so they indirectly
induce an accelerated decrease of $m_{H_{2}}^{2}$ compared to
$m_{H_{1}}^{2}$. On the other hand large $m_{0}$ values help
$m_{H_{1}}^{2}$ decrease faster than $m_{H_{2}}^{2}$, mainly because of the
presence of the $\tau$--Yukawa term in the equation for $m_{H_{1}}^{2}$.
Since both effects are more or less equally significant, an appropriate
difference between $m_{H_{1}}^{2}$, $m_{H_{2}}^{2}$
can only be achieved if $M_{1/2}$
is larger than ${m_{0}}$.
Furthermore there has to be a lower bound on $M_{1/2}$, since a
minimum value of $M_{1/2}$ is needed to make $m_{U}^{2}$ sufficiently
larger than $m_{D}^{2}$, or indirectly $m_{H_{2}}^{2}$ sufficiently
smaller than $m_{H_{1}}^{2}$.

Returning back to Fig. 1 then, one observes that disentangling the Higgs
mass parameter $m_{H}$ from $m_{0}$ (case B), does not significantly
change the universal picture: $m_{0}$ and $m_{H}$ must again be small
for the $\tau$--term not to contribute significantly to the running of
the $m_{H_{i}}$'s, $M_{1/2}$ must still be large to induce the right
$m_{U}^{2}$--$m_{D}^{2}$ splitting which will lift the degeneracy of the
Higgs doublets at low energies.
If however this degeneracy
is already lifted at the unification scale (case C), even by a fairly
mild 20$\%$, the solution space in the $m_{0}$--$M_{1/2}$ plane
increases drastically:
$m_{0}$ need no longer be smaller than $M_{1/2}$ and  $M_{1/2}$ is restricted
 from below only through the experimental bound on the gluino.
These two effects have important consequences for the predicted
sparticle spectrum and therefore also for the supersymmetric corrections to
the bottom mass as we will see in the discussion of Fig. 2 and 4.
In all three cases A,B and C
very small values of $m_{0}$ would result in the lightest stau
being lighter than the lightest neutralino and we exclude these solutions
by requiring the lightest supersymmetric particle (LSP) to carry no charge.

In Fig. 2 the $M_{1/2}$--$\mu$ plane is depicted with $\mu$ evaluated at
$M_{Z}$. The tight correlation between $M_{1/2}$ and $\mu$ in the universal
case has already been pointed out in ref. \cite{copw2}, where it has also been
described with the help of semi--analytical formulas. A crude way to understand
this correlation, at least for most of the solutions, is the following:
At low energies both $m_{H_{1}}^{2}$ and $m_{H_{2}}^{2}$
are
negative\footnote{The Yukawa terms in their RGE's dominate over the gauge
terms.},
with $m_{H_{2}}^{2}$ smaller than $m_{H_{1}}^{2}$. For a proper radiative
breakdown $\mu$ must then be such that $m_{1}^{2}>0$ while $m_{2}^{2}<0$
($m_{i}^{2}=m_{H_{i}}^{2}+\mu^{2}$). In other words $\mu^{2}$=${\cal O}\left(-
\frac{m_{H_{1}}^{2}+m_{H_{2}}^{2}}{2}\right)$ ( the difference of
$m_{H_{1}}^{2}$, $m_{H_{2}}^{2}$ is in general much smaller than their
absolute values). Now raising $M_{1/2}$ has the effect of
driving $m_{H_{1}}^{2}$, $m_{H_{2}}^{2}$ to lower
values\footnote{Larger $M_{1/2}$ means heavier squarks and therefore
larger
contributions from the Yukawa terms. The gauge terms do not compensate
these contributions because they contain only the electroweak sector.},
 so we need larger
values of $\mu$ to fulfill the symmetry breaking conditions. The dependence
 of $\mu$ on $M_{1/2}$ turns out to be practically linear in the universal
case for the parameter space scanned. The lower limit on $M_{1/2}$
has been explained in the discussion of Fig. 1. Fig. 2 confirms what
has already been observed in Fig. 1: The effect of treating the
Higgses independently (case B) is minimal, preserving, though slightly
loosening the linear $M_{1/2}$--$\mu$ correlation. However, when one lifts
the mass--degeneracy of the Higgses at $M_{X}$ (case C), a much larger
area of the $\mu$--$M_{1/2}$ plane gives proper radiative breakdown. The
correlation between $\mu$ and $M_{1/2}$ is in this case still valid,
since it remains true that raising $M_{1/2}$ drives the $m_{H_{i}}^{2}$'s
smaller, thus requiring larger $\mu$ for proper symmetry breaking. However,
since in this case also large values of $m_{0}$ are permitted, $m_{0}$
contributes significantly in the running of the $m_{H_{i}}^{2}$'s when
$M_{1/2}$ is small
(it is $m_{0}$ that makes the squarks heavy when $M_{1/2}$ is small).
For this reason the dependence of $\mu$ on $M_{1/2}$ clearly
deviates from linearity in the lower  part of the $\mu$--$M_{1/2}$ plane.
This has
the effect that even for very small values of $M_{1/2}$, $\mu$ still remains
fairly large.

Let us now turn to some characteristic features of the predicted
sparticle spectrum. An important observation is that radiative electroweak
breaking requires $\mu$ to be always significantly
 larger than $M_{1/2}$. This has
the effect that the mixing of the neutralinos and the charginos is
small and therefore the lightest neutralino and the lightest chargino
are an almost pure bino and an almost pure wino respectively. In Fig. 3
we plot the lightest chargino tree--level mass versus $\mu$ for the
three cases A,B and C. A noteworthy feature is that certain
non--universalities (e.g. case C) allow for fairly light chargino
masses, whereas with universal soft terms there is a lower bound of
$\sim$300GeV for the lightest chargino. The same applies for the lightest
neutralino which can be as light as $\sim$45GeV in case C while in the
universal case it is always heavier than $\sim$150GeV. These observations
are of course directly related to the lower bound on $M_{1/2}$ in the
various cases.

Closely related to the predicted sparticle spectrum are the
corrections to the bottom mass coming from sparticle loops. The dominant
corrections come from sbottom--gluino and stop--chargino loops \cite{hrs1}.
Defining  the running bottom mass by $m_{b}=h_{b}\upsilon_{1}(1+\delta
m_{b})$, the magnitude of the corrections is given
by the approximate
formula \cite{copw2}:
\ba
\delta m_{b} &=& \frac{2\alpha_{3}}{3\pi}K_{1}tan\beta\frac{M_{\tilde{g}}\mu}
{m_{max1}^{2}}\nonum
&&{}+\frac{h_{t}^{2}}{16\pi^{2}}K_{2}tan\beta\frac{A_{t}\mu}
{m_{max2}^{2}}\ ,
\label{eq:cor}
\ea
where $K_{1}$, $K_{2}$ are coefficients of order one, $M_{\tilde{g}}$ is the
gluino mass and $m_{max1}^{2}$, $m_{max2}^{2}$ the squared masses of the
heaviest particles running in the corresponding loops. We evaluate the
corrections at the electroweak scale. Both corrections are proportional to
\tb and therefore can be very large if the mass ratios appearing in
(\ref{eq:cor}) are not small. It is clear from Fig. 2 that \mg and $\mu$
are closely correlated ($M_{\tilde{g}}$=$\frac{\alpha_{3}}{\alpha_{G}}M_{1/2}
\simeq{3M_{1/2}}$ at $M_{Z}$). Also since the running of $A_{t}$ is very
similar to that of $M_{\tilde{g}}$, $A_{t}$ is always roughly of the order
of $M_{\tilde{g}}$. In the universal scenario \mg is very heavy
because of the lower limit on $M_{1/2}$, so all three quantities
$M_{\tilde{g}}$, $\mu$ and $A_{t}$ are large. The heaviest particles running
in the two relevant loops are in  this scenario the gluino and the
heavy stop respectively. Since however also the heavy squarks are of the
order of the
gluino mass, the ratios
appearing in (\ref{eq:cor}) are ${\cal O}$(1)
and the corrections to the bottom mass are not suppressed.
The situation is the same if one disentangles the Higgses from squarks and
sleptons, since the predicted spectrum is very much like the one of the
universal scenario. The possibility of suppressing the corrections
arises if one considers case C. The fact that very
low values for $M_{1/2}$ are allowed in this case, while at the same
time $m_{0}$ can be large (see Fig. 1), gives solutions with squarks quite
heavier than $M_{\tilde{g}}$ and $A_{t}$. This means small ratios in
(\ref{eq:cor}) and therefore suppressed corrections.

The above remarks are reflected in Fig. 4. For the cases A and B
the supersymmetric corrections to the bottom mass are in all solutions
of the order of 30\%, positive or negative depending on the sign
of $\mu$. So in these cases the prediction $M_{b}$=5.5GeV for the
uncorrected mass is modified to $M_{b}$=4.2GeV or $M_{b}$=6.9GeV
(depending on the sign) when corrections are taken into account.
Both value lie outside the range quoted for the b pole--mass. In case
C the situation is different: There are solutions covering the
whole range of corrections between 5\% and 60\%. So although also in this
case very large corrections are not excluded, the solutions with
$M_{1/2}$ small provide the possibility of suppressing them and
successfully predicting the bottom quark mass: Negative corrections
between 15\% and 5\% would give 4.7GeV$\le{}M_{b}\le{}$5.2GeV.

Until now we have concentrated on non-universalities directly related
to the Higgs sector, a natural thing to do if one wants to study
the radiative breaking of the electroweak symmetry. However non-universalities
introduced in other sectors of the theory can also affect the process of
symmetry breaking, since these sectors couple to the Higgs sector and
therefore
affect the running of the Higgs potential parameters. Relaxing the universality
within the gaugino sector at $M_{X}$ causes negligible modifications of the
solution space. This is so because essentially only $M_{1}$ plays a decisive
role in the process of symmetry breaking (it differentiates the running of
$m_{U}^{2}$, $m_{D}^{2}$ and thus also that of $m_{H_{i}}^{2}$). So if the
$M_{i}$'s are not universal at $M_{X}$ then  $M_{1}$($M_{X}$) plays the role
that $M_{1/2}$ plays in the universal case, at least as far as the symmetry
breaking is concerned.
However, lifting the universality
between up-- and down-- type squarks at $M_{X}$ has relevant
effects\footnote{We thank S.Pokorski for discussions on this point. For a
thorough investigation see ref. \cite{op2}.},
since
$m_{U}^{2}$ and $m_{D}^{2}$ induce a non--uniform evolution of the
$m_{H_{i}}^2$'s, as has previously been explained. This is an indirect way
to make the two Higgses non-degenerate, a fact also reflected in the
solution space of such scenarios which looks very similar to that of case C.
It should however be noted that because of the indirect way the non--degeneracy
is induced, a mass splitting of the order of 40\% between up-- and down--type
squarks is needed, in order to significantly affect the process of
radiative electroweak breaking. Mass splittings of the order of 20\% only
reproduce the solution space of the universal scenario.

Let us finally turn to the low \tb regime. We will present illustrative
 results
for tan$\beta$=1.8, for which unification predicts the strong gauge
coupling
to be $\alpha_3(M_{Z})$=0.129 and the top pole--mass $M_{t}$=187GeV,
when $T_{SUSY}=M_{Z}$ (for $T_{SUSY}$=500GeV one finds
$\alpha_3(M_{Z})\simeq$0.123 and
$M_{t}\simeq$185GeV).

In Fig. 5 we show the $m_{0}$--$\mu$ plane with the three types of regions
corresponding again to the cases A,B and C defined previously.
Let us first consider more closely the universal scenario. When \tb
is small there is a large hierarchy between $h_{t}$ and $h_{b}$
(and between $h_{t}$ and $h_{\tau}$, of course,
but we ignore the latter in the following discussion since its
contribution is very small). Moreover the requirement of b--$\tau$
unification is only fulfilled for very large values of $h_{t}$
\cite{a1,cpw,lp2},
which means that at low energies $h_{t}$ is always very close to its
infrared quasi--fixed point value \cite{fp}. These two facts make
$h_{t}$ dominate in the process of radiative electroweak
symmetry breaking when \tb is small. This has as a consequence that if
one wants to drive $m_{H_{2}}^{2}$ smaller than $m_{H_{1}}^{2}$
both $M_{1/2}$ and $m_{0}$ contribute in the right direction.
For $M_{1/2}$ the reasons given in the large \tb discussion apply
here as well (it makes $m_{U}^{2}>m_{D}^{2}$)\footnote{In fact, since
for small \tb the Yukawa terms in the equation for
$m_{H_{1}}^{2}$ are negligible, the splitting of $m_{U}^{2}$, $m_{D}^{2}$
is no longer crucial: making squarks heavy is enough to accelerate the
decrease of $m_{H_{2}}^{2}$ and this is the actual role of $M_{1/2}$ in
the low \tb regime.}.
Now however also $m_{0}$
tends to make $m_{H_{2}}^{2}$ smaller than $m_{H_{1}}^{2}$ and
not vice versa as was the case for large tan$\beta$.
This is again the effect of the Yukawa terms
and the fact that $h_{b}$, $h_{\tau}$ are now
negligible compared to $h_{t}$. One further important feature when
\tb is small is that large values of $M_{1/2}$ drive certain squarks
very heavy. This is so because  of the $h_{b}$ term in the $m_{D}^{2}$
equation, which tends to decrease the value of $m_{D}$,
has a vanishing contribution. Therefore, by setting an upper bound
of 2.5TeV on the squark masses, we effectively forbid large values for
the $M_{1/2}$ parameter, in particular when $m_{0}$ is large.
This means that for most of the solutions $m_{0}$ is larger than $M_{1/2}$
and dominates in the process of symmetry breaking. This explains the strong
correlation between $m_{0}$ and $\mu$. Also the lower bound on $m_{0}$,
present in Fig. 5 in the universal case, is the effect of setting an
upper bound of 2.5TeV on the squark masses: If $m_{0}$ were to be
vanishingly small, a fairly large value of $M_{1/2}$ would be
needed to induce the splitting of the Higgs doublets. Such a value
would however drive certain squarks unacceptably heavy, and this is
the reason why solutions with very small $m_{0}$ have been excluded.
One should however note that if very heavy squarks were not considered to be a
problem, symmetry breaking alone does not forbid vanishing $m_{0}$.

Returning then to Fig. 5 let us see how non--universalities affect
the above picture. It is now case C (non--degenerate Higgses at $M_{X}$)
that has negligible effects on the universal results: The hierarchy of the
$m_{H_{i}}^{2}$'s can be easily induced radiatively, non--universal
initial conditions do not enlarge the solution space. However, if one
reserves $m_{0}$ for squarks and sleptons only and treats the Higgses
independently (case B), there are observable effects in the $m_{0}$--$\mu$
plane: Solutions with very small values of $m_{0}$ are allowed. The reason for
this is that $m_{H}$ can in this case compensate for the vanishing contribution
of $m_{0}$ in the running of the $m_{H_{i}}^{2}$'s, so $M_{1/2}$ does not
have to be so large that squarks become too heavy. In short we can
have vanishing $m_{0}$ with squarks below 2.5TeV.

\section{CONCLUSIONS AND OUTLOOK}

We thus have seen that many predictions of conventional SUSY--GUT's
are changed in the presence of non--universal soft breaking terms.
The importance of this non--universality is most prominent in
SO(10)--like models with large tan$\beta$. There some of the main
conclusions of the universal case have to be revised. Small values of
$M_{1/2}$ are now allowed bringing back the lightest charginos
and neutralinos in the experimentally accessible mass range. The presence
of these light particles in the large \tb regime might also have
consequences for a discussion of the supersymmetric corrections to the
Z$b\overline{b}$--vertex.
The presence of non--universal terms may also be important for the size of
radiative corrections to the b--quark mass from supersymmetric
particles. While in the universal case these corrections are always
large (when \tb is large), they might be reduced in the non-- universal case.

Fortunately, however, there are many situations where the effects
of non-universal terms are less important. In the presence of large
$M_{1/2}$ this happens very often since in that case the radiative
corrections to the soft masses tend to wash out any primeval
non--universality. The case of small \tb is less affected by
non--universality, as could have been expected. Here a split of the
two Higgs masses is not so important, since radiative symmetry
breakdown can be achieved for a wide region of the (universal) parameter
space. On the other hand a situation where the Higgs masses are split
 from squark and slepton masses, leads to the new possibility of small
$m_{0}$, excluded in the universal case. Otherwise the small \tb
results for the universal case are pretty stable.
Even in the case of large \tb some predictions remain rather insensitive
to non--universalities. In particular we note the the stability of the
$\mu$--$M_{1/2}$ correlation as shown in Fig. 2.

Thus many results of the universal case might have more general
validity. Nonetheless one should be aware of the fact, that in a given
model the specific predictions should be carefully worked out.
Especially in models with large \tb we find large sensitivity to a
non--universality of the soft terms that might have decisive
influence on the phenomenological properties. Let us all hope that the
results of LEP II will shed some light on these questions and help us
to determine the values of the parameters

 ~\\~\\

\end{document}